\documentclass[a4paper,11pt]{article}
\usepackage[top=1.2in,right=1in,left=1in,bottom=1.2in]{geometry}
\usepackage{latexsym}
\usepackage{amssymb}
\usepackage{graphics,graphicx}
\usepackage{tabularx}
\usepackage{amsfonts}
\usepackage{amsmath}
\usepackage{subfig}
\usepackage[T1]{fontenc} 
\usepackage[latin1]{inputenc} 
\usepackage{epstopdf}
\usepackage{stix}
\usepackage[normalem]{ulem}
\usepackage{latexsym}
\usepackage{amssymb}
\usepackage{float}
\usepackage{tabularx}
\usepackage{amsfonts}
\usepackage{amsmath}
\usepackage{bm}
\usepackage{empheq}
\usepackage{enumerate}
\usepackage{booktabs}
\usepackage{multirow}
\usepackage{relsize}
\usepackage{subfig}
\usepackage{color}
\usepackage{mathtools}
\usepackage{bmpsize}
\usepackage{xcolor}
\usepackage{siunitx}
\usepackage{graphicx}

\usepackage{color}

\newcommand{\footmsg}[1]{%
  \let\temp\thempfn%
  \def\thempfs{}
  \footnotetext{#1}
  \let\tempfn\temp}

\usepackage{fancyhdr}            
\fancypagestyle{plain}{%
	\fancyhead{}                                     
	\fancyfoot[CE,CO]{\thepage}       
	\fancyhead[CE,CO]{\textcolor[rgb]{0.64,0.15,0.15}{Published in \emph{Journal of Applied Mechanics} (2021), 88(3): 031003 
			\\
			doi:  https://doi.org/10.1115/1.4047132}}
	\setlength{\headheight}{14.5pt}}

\begin{document}

\newcommand{\singlespace} {\baselineskip=12pt
\lineskiplimit=0pt \lineskip=0pt }
\def\ds{\displaystyle}

\newcommand{\beq}{\begin{equation}}
\newcommand{\eeq}{\end{equation}}
\newcommand{\lb}{\label}
\newcommand{\beqar}{\begin{eqnarray}}
\newcommand{\eeqar}{\end{eqnarray}}
\newcommand{\barr}{\begin{array}}
\newcommand{\earr}{\end{array}}

\newcommand{\jump}{\parallel}

\def\c{{\circ}}

\newcommand{\Ehat}{\hat{E}}
\newcommand{\That}{\hat{\bf T}}
\newcommand{\Ahat}{\hat{A}}
\newcommand{\chat}{\hat{c}}
\newcommand{\shat}{\hat{s}}
\newcommand{\khat}{\hat{k}}
\newcommand{\muhat}{\hat{\mu}}
\newcommand{\mc}{M^{\scriptscriptstyle C}}
\newcommand{\mei}{M^{\scriptscriptstyle M,EI}}
\newcommand{\mec}{M^{\scriptscriptstyle M,EC}}

\newcommand{\hbeta}{{\hat{\beta}}}
\newcommand{\rec}[2]{\left( #1 #2 \ds{\frac{1}{#1}}\right)}
\newcommand{\rep}[2]{\left( {#1}^2 #2 \ds{\frac{1}{{#1}^2}}\right)}
\newcommand{\derp}[2]{\ds{\frac {\partial #1}{\partial #2}}}
\newcommand{\derpn}[3]{\ds{\frac {\partial^{#3}#1}{\partial #2^{#3}}}}
\newcommand{\dert}[2]{\ds{\frac {d #1}{d #2}}}
\newcommand{\dertn}[3]{\ds{\frac {d^{#3} #1}{d #2^{#3}}}}

\def\scalp{\mbox{\boldmath$\, \cdot \, $}}
\def\gdp{\makebox{\raisebox{-.215ex}{$\Box$}\hspace{-.778em}$\times$}}

\def\daa{\makebox{\raisebox{-.050ex}{$-$}\hspace{-.550em}$: ~$}}

\def\mK{\mbox{${\mathcal{K}}$}}
\def\cK{\mbox{${\mathbb {K}}$}}

\def\Xint#1{\mathchoice
   {\XXint\displaystyle\textstyle{#1}}%
   {\XXint\textstyle\scriptstyle{#1}}%
   {\XXint\scriptstyle\scriptscriptstyle{#1}}%
   {\XXint\scriptscriptstyle\scriptscriptstyle{#1}}%
   \!\int}
\def\XXint#1#2#3{{\setbox0=\hbox{$#1{#2#3}{\int}$}
     \vcenter{\hbox{$#2#3$}}\kern-.5\wd0}}
\def\ddashint{\Xint=}
\def\fpint{\Xint=}
\def\dashint{\Xint-}
\def\cpvint{\Xint-}
\def\intl{\int\limits}
\def\cpvintl{\cpvint\limits}
\def\fpintl{\fpint\limits}
\def\ointl{\oint\limits}

\def\bA{{\bf A}}
\def\ba{{\bf a}}
\def\bB{{\bf B}}
\def\bb{{\bf b}}
\def\bc{{\bf c}}
\def\bC{{\bf C}}
\def\bD{{\bf D}}
\def\bd{{\bf d}}
\def\bE{{\bf E}}
\def\be{{\bf e}}
\def\bbf{{\bf f}}
\def\bF{{\bf F}}
\def\bG{{\bf G}}
\def\bg{{\bf g}}
\def\bi{{\bf i}}
\def\bH{{\bf H}}
\def\bK{{\bf K}}
\def\bL{{\bf L}}
\def\bM{{\bf M}}
\def\bN{{\bf N}}
\def\bn{{\bf n}}
\def\bm{{\bf m}}
\def\b0{{\bf 0}}
\def\bo{{\bf o}}
\def\bX{{\bf X}}
\def\bx{{\bf x}}
\def\bP{{\bf P}}
\def\bp{{\bf p}}
\def\bQ{{\bf Q}}
\def\bq{{\bf q}}
\def\bR{{\bf R}}
\def\bS{{\bf S}}
\def\bs{{\bf s}}
\def\bT{{\bf T}}
\def\bt{{\bf t}}
\def\bU{{\bf U}}
\def\bu{{\bf u}}
\def\bv{{\bf v}}
\def\bw{{\bf w}}
\def\bV{{\bf V}}
\def\bW{{\bf W}}
\def\by{{\bf y}}
\def\bz{{\bf z}}
\def\bk{{\bf k}}

\def\Id{{\bf I}}
\def\bxi{\mbox{\boldmath${\xi}$}}
\def\balpha{\mbox{\boldmath${\alpha}$}}
\def\bbeta{\mbox{\boldmath${\beta}$}}
\def\bchi{\mbox{\boldmath${\chi}$}}
\def\bepsilon{\mbox{\boldmath${\epsilon}$}}
\def\bvarepsilon{\mbox{\boldmath${\varepsilon}$}}
\def\bomega{\mbox{\boldmath${\omega}$}}
\def\bphi{\mbox{\boldmath${\phi}$}}
\def\bsigma{\mbox{\boldmath${\sigma}$}}
\def\bfeta{\mbox{\boldmath${\eta}$}}
\def\bDelta{\mbox{\boldmath${\Delta}$}}
\def\btau{\mbox{\boldmath $\tau$}}

\def\tr{{\rm tr}}
\def\dev{{\rm dev}}
\def\div{{\rm div}}
\def\Div{{\rm Div}}
\def\Grad{{\rm Grad}}
\def\grad{{\rm grad}}
\def\Lin{{\rm Lin}}
\def\Sym{{\rm Sym}}
\def\Skw{{\rm Skew}}
\def\abs{{\rm abs}}
\def\Re{{\rm Re}}
\def\Im{{\rm Im}}

\def\forE{\mathbb E}
\def\forK{\mathbb K}
\def\capB{\mbox{\boldmath${\mathsf B}$}}
\def\capC{\mbox{\boldmath${\mathsf C}$}}
\def\capD{\mbox{\boldmath${\mathsf D}$}}
\def\capE{\mbox{\boldmath${\mathsf E}$}}
\def\capG{\mbox{\boldmath${\mathsf G}$}}
\def\tcapG{\tilde{\capG}}
\def\capH{\mbox{\boldmath${\mathsf H}$}}
\def\capK{\mbox{\boldmath${\mathsf K}$}}
\def\capL{\mbox{\boldmath${\mathsf L}$}}
\def\capM{\mbox{\boldmath${\mathsf M}$}}
\def\capR{\mbox{\boldmath${\mathsf R}$}}
\def\capW{\mbox{\boldmath${\mathsf W}$}}

\def\i{\mbox{${\mathrm i}$}}

\def\mC{\mbox{\boldmath${\mathcal C}$}}

\def\mB{\mbox{${\mathcal B}$}}
\def\mE{\mbox{${\mathcal{E}}$}}
\def\mL{\mbox{${\mathcal{L}}$}}
\def\mK{\mbox{${\mathcal{K}}$}}
\def\mV{\mbox{${\mathcal{V}}$}}

\def\C{\mbox{\boldmath${\mathcal C}$}}
\def\E{\mbox{\boldmath${\mathcal E}$}}

\def\ARMA{{ Arch. Rat. Mech. Analysis\ }}
\def\AMR{{ Appl. Mech. Rev.\ }}
\def\ASCEEM{{ ASCE J. Eng. Mech.\ }}
\def\acta{{ Acta Mater. \ }}
\def\CMAME {{ Comput. Meth. Appl. Mech. Engrg.\ }}
\def\CRAS{{ C. R. Acad. Sci., Paris\ }}
\def\EFM{{ Eng. Fracture Mechanics\ }}
\def\EJMA{{ Eur.~J.~Mechanics-A/Solids\ }}
\def\IJES{{ Int. J. Eng. Sci.\ }}
\def\IJF{{ Int. J. Frac.\ }}
\def\IJMS{{ Int. J. Mech. Sci.\ }}
\def\IJNME{{ Int. J. Num. Meth. Engng.\ }}
\def\IJNAMG{{ Int. J. Numer. Anal. Meth. Geomech.\ }}
\def\IJP{{ Int. J. Plasticity\ }}
\def\IJSS{{ Int. J. Solids Structures\ }}
\def\IngA{{ Ing. Archiv\ }}
\def\JAM{{ J. Appl. Mech.\ }}
\def\JAP{{ J. Appl. Phys.\ }}
\def\JE{{ J. Elasticity\ }}
\def\JM{{ J. de M\'ecanique\ }}
\def\JMM{{ J. Micromech. Microeng.\ }}
\def\JMPS{{ J. Mech. Phys. Solids\ }}
\def\JOMMS{{ J. Mech. Materials Struct.\ }}
\def\Macro{{ Macromolecules\ }}
\def\MOM{{ Mech. Materials\ }}
\def\MMS{{ Math. Mech. Solids\ }}
\def\MPCPS{{ Math. Proc. Camb. Phil. Soc.\ }}
\def\MRC{{ Mech. Res. Comm.\ }}
\def\MSE{{ Mater. Sci. Eng.}}
\def\nature{{ Nature\ }}
\def\PM{{Phil. Mag.\ }}
\def\PMPS{{ Proc. Math. Phys. Soc.\ }}
\def\PRSA{{ Proc. R. Soc. A\ }}
\def\PRSL{{ Proc. R. Soc.\ }}
\def\rock{{ Rock Mech. and Rock Eng.\ }}
\def\QAM{{ Quart. Appl. Math.\ }}
\def\QJMAM{{ Quart. J. Mech. Appl. Math.\ }}


\def\salto#1#2{
\left[\mbox{\hspace{-#1em}}\left[#2\right]\mbox{\hspace{-#1em}}\right]}




\title{Flutter instability and Ziegler destabilization paradox\\ for elastic rods subject to non-holonomic constraints}

\author{Alessandro Cazzolli, Francesco Dal Corso, Davide Bigoni$^{*}$\\ 
DICAM, University of Trento, via Mesiano 77, I-38123 Trento, Italy\\
e-mail: francesco.dalcorso@unitn.it,  davide.bigoni@unitn.it\\
}
\date{}
\maketitle
\footnotetext[1]{Corresponding author: Davide Bigoni 
fax: +39 0461 282599; tel.: +39 0461 282507; web-site:
http://bigoni.dicam.unitn.it/; e-mail:
davide.bigoni@unitn.it}

\begin{abstract} 

Two types of non-holonomic constraints (imposing a prescription on velocity) are analyzed, connected to an end of a (visco)elastic rod, straight in its undeformed configuration. 
The equations governing the nonlinear dynamics are obtained and then linearized near the trivial equilibrium configuration. 
The two constraints are shown to lead to the same equations governing the linearized dynamics of the Beck (or Pfl\"uger) column in one case and of the Reut column in the other.
Therefore, although the structural systems are fully conservative (when viscosity is set to zero), they exhibit flutter and divergence instability. In addition, the Ziegler's destabilization paradox is found when dissipation sources are introduced. It follows that these features are proven to be not only a consequence of  \lq unrealistic non-conservative loads' (as often stated in the literature), rather, the models proposed by Beck, Reut, and Ziegler can exactly describe the linearized dynamics of structures subject to non-holonomic constraints, which are made now fully accessible to experiments.
\end{abstract}

\section{Introduction}

\begin{figure}[!htb]
	\begin{center}
		\includegraphics[width=3.in]{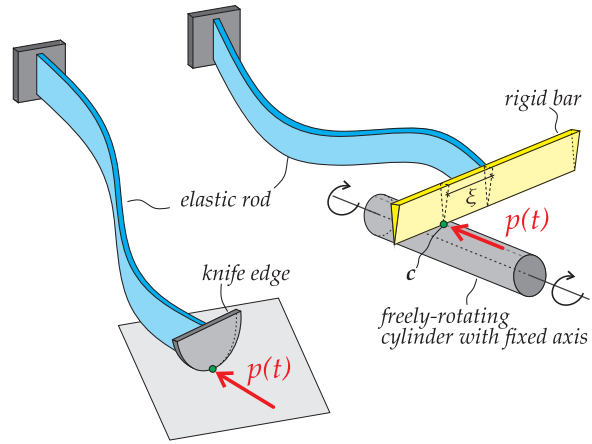}
	\end{center}
	\caption{\footnotesize{Two types of the non-holonomic constraints applied to an elastic rod. The \lq skate' constraint (left) realized through a knife edge (a non-slipping wheel can also be used), and the \lq violin bow' constraint (right) realized through slipless contact between a cylinder, freely rolling about its fixed axis, and a rigid bar attached orthogonally to the end of the rod.}}
	\label{fig0}
\end{figure}

Elastic structures subject to non-holonomic constraints, such as a rolling wheel or sphere, do not admit a formulation in terms of energy potential, even when all applied loads are conservative \cite{neimark,leipholz}. For these mechanical systems, the linearized equations governing the dynamics can be 
characterized by a non-hermitian operator \cite{ruina,bottema} admitting complex conjugate eigenvalues, so that flutter instabilities may occur. An example of this behaviour is provided by 
the so-called \lq shimmy instability' \cite{beregi1,ziegshimmy,ziegler}.  The scope of the present article is an extension of results obtained for discrete systems \cite{cazzollinhol} to continuous structures equipped with non-holonomic constraints. In particular, two types of non-holonomic constraints attached to an elastic (or visco-elastic) rod are investigated, as sketched in Fig.\ref{fig0}.
The first type (referred as \lq skate', Fig.\ref{fig0} left) can be visualized as a knife edge or a non-slipping wheel mounted at the end of a rod with an inclination $\beta_0$ (which is detailed in Fig. \ref{fig_cont}). The reaction $\boldsymbol{p}$ (of modulus $p$) transmitted to the structure by this device is directed along the \lq skate' axis 
\begin{equation}
\label{numerouno1}
\boldsymbol{p} = \boldsymbol{e}_1 p\,\cos\left(\theta(l) + \beta_0\right) + \boldsymbol{e}_2 p\,\sin\left(\theta(l) + \beta_0\right), 
\end{equation}
where $\theta(l)$ is the rotation of the rod's end and $ \boldsymbol{e}_1$ and $ \boldsymbol{e}_2$ are two orthogonal unit vectors providing the reference system. 
The key issue associated to the non-holonomic \lq skate' constraint is that the {\it velocity} of the end of the structure $\boldsymbol{v}(l)$ is prescribed to remain orthogonal to $\boldsymbol{p}$
\begin{equation}
\label{numerodue1}
\boldsymbol{p} \cdot \boldsymbol{v}(l) = 0.
\end{equation}
The other non-holonomic constraint (referred as \lq violin bow', Fig.\ref{fig0} right) investigated here is a device in which a rigid \lq appendix' of an elastic structure is constrained to remain in slipless contact with a circular cylinder (inclined at an angle $\beta_0$ detailed in Fig. \ref{fig_cont}) which can only roll about its axis. In this case the reaction on the structure remains coaxial with the cylinder
\begin{equation}
\boldsymbol{p}= \boldsymbol{e}_1 p\,\cos\beta_0 + \boldsymbol{e}_2 p\,\sin\beta_0 .
\end{equation}
Moreover, the velocity $\dot{\boldsymbol{c}}$ of the point $\boldsymbol{c}$ belonging to the rigid appendix of the structure in contact with the cylinder (Fig.\ref{fig0}, right) is given by the Poisson's theorem as 
\begin{equation}
\dot{\boldsymbol{c}} = \boldsymbol{v}(l) + \dot{\theta}(l) \boldsymbol{e}_3 \times \xi \left(-\boldsymbol{e}_1\sin\theta(l) + \boldsymbol{e}_2\cos\theta(l) \right) 
\end{equation}
(where $\xi$ is the distance along the rigid appendix to the connection with the elastic rod and $\boldsymbol{e}_3 = \boldsymbol{e}_1\times\boldsymbol{e}_2$) and, to satisfy the non-holonomic \lq violin bow' constraint, has to remain orthogonal to the reaction $\boldsymbol{p}$
\begin{equation}
\label{numerodue2}
\boldsymbol{p} \cdot \dot{\boldsymbol{c}} = 0.
\end{equation}

It is clear from the above discussion that the essence of non-holonomy is a requirement on the {\it velocity} (not displacement!) at some point of the structure. This leads to the fact that the reaction $\boldsymbol{p}$ does zero work during every possible motion and therefore {\it the mechanical system is conservative} if the structure is purely elastic and subject to a load admitting a potential \cite{neimark,ruina}. 

Subject to two different conservative loadings (that will be detailed later), {\it the linearized dynamic equations for the rod are derived in this article\footnote{The behaviour of the elastic rod analyzed in this paper is also confirmed using a discretized rod model \cite{cazzollinhol} at increasing number of its constituent rigid bars, so that the continuous and discrete models show coincidence of the instability loads in the limit.}and 
	are shown to coincide with the corresponding equations holding for the Beck and Pfl\"uger columns \cite{beck, bigonimiss22, bigonikiri, detinko, pfluger}, when the \lq skate' constraint is applied, or to the Reut column \cite{bolotin, reut}, when the \lq violin bow' is considered}. Therefore, the structures exhibit {\it flutter and divergence instability},  although the system (in the absence of viscous dissipation) is fully conservative. Moreover, when viscosity is present, the Ziegler destabilization paradox \cite{ bigonikiri ,cazzollinhol ,kirillov_1 ,kirillov_2 ,ziegler} occurs in the limit of vanishing dissipation. 

Research on structures subject to nonconservative loads is a timely topic in view of several, different applications \cite{abdulla, agostinelli, phan}, nevertheless the fundamental models proposed by Reut, Ziegler, and Beck have often been considered unrealistic and therefore harshly criticized \cite{koiter, elishakoff}. The results presented here (and those relative to discrete systems \cite{cazzollinhol}) demonstrate that {\it the Reut, Ziegler, and Beck models represent the exact linearized behaviour of corresponding structures subject to non-holonomic constraints}.

Finally, the obtained results confirm the validity of the experimental verification of the flutter load proposed in \cite{bigoninoselli, bigonimiss22, bigonikiri, bigmiss} and 
definitely open the way to exploitation of non-holonomic constraints in the mechanics of deformable structures. 

\section{Viscoelastic rod subject to a non-holonomic constraint}

\begin{figure}[!htb]
	\begin{center}
		\includegraphics[width=3.3in]{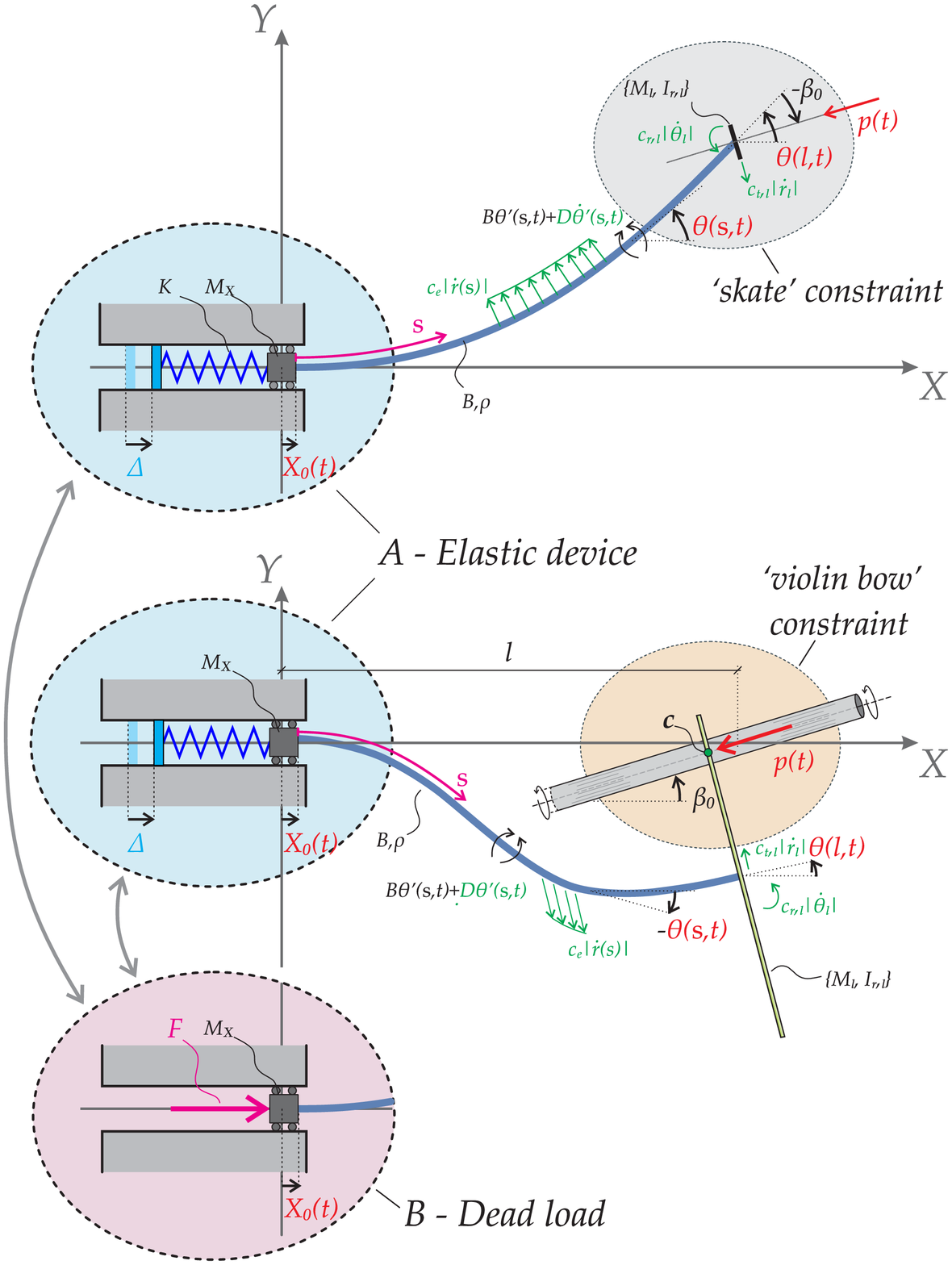}
	\end{center}
	\caption{\footnotesize{A flexible rod (bending stiffness $B$ and mass density $\rho$) is subject to either the non-holonomic \lq skate' or \lq violin bow' constraint. The rod's deformed configuration is defined by the rotation $\theta(\textsf{s},t)$ and the horizontal position $X_0(t)$ of a sliding clamp of mass $M_X$ which is loaded either with a precompressed linear spring (of stiffness $K$) or with a dead load $F$.}}
	\label{fig_cont}
\end{figure}

An inextensible and unshearable rod of length $l$ is considered in the plane $X-Y$, straight in its  underformed configuration.  The rod's tangent is inclined (with respect to the $X$-axis) at an angle $\theta(\textsf{s},t)$, function of the curvilinear coordinate $\textsf{s}\in[0,l]$ and time $t$,  Fig.\ref{fig_cont}. The rod's end (at $\textsf{s}=0$) is constrained with a clamp of mass $M_X$, which can freely slide along the $X-$axis, so that $\theta(0,t)=0$ and its position is singled out by the coordinate $X_0(t)=X(\textsf{s}=0,t)$.
In this setting, the usual expressions of the deformed shape of an inextensible rod hold 
\begin{equation}
\label{kinem}
\begin{split}
X(\textsf{s},t)&=X_0(t)+\int_0^\textsf{s} \cos \theta(\varsigma,t) \text{d}\varsigma,\\ Y(\textsf{s},t)&=\int_0^\textsf{s} \sin \theta(\varsigma,t) \text{d}\varsigma,
\end{split}
\end{equation}
showing that the rod's motion is described by $\theta(\textsf{s},t)$ and the clamp position $X_0(t)$. 

A conservative loading is applied to the sliding clamp by means of (Fig.\ref{fig_cont})
\begin{itemize}
	\item \textit{A - a linear elastic spring} of stiffness $K$, initially pre-compressed by a prescribed displacement $\Delta>0$ (fixed in time), so that its elastic energy is given by
	\begin{equation}
	\label{plof1}
	\Xi(X_0)= \dfrac{1}{2} K \left[X_0-\Delta\right]^2 ;
	\end{equation}
	
	\item \textit{B - a dead load} $F$, with potential energy 
	\begin{equation}
	\label{plof2}
	\Xi(X_0)=-F\, X_0 . 
	\end{equation}
	
\end{itemize}
The two types of non-holonomic constraint mentioned in the introduction are applied at the right rod's end ($\textsf{s}=l$):
\begin{itemize}
	
	\item \textit{S - the \lq skate' constraint}, expressed by equation (\ref{numerodue1}), which is rewritten as 
	\begin{equation}
	\label{skateconstraint}
	\dot X(l)\cos\left[\theta(l)+\beta_0\right]+\dot Y(l)\sin\left[\theta(l)+\beta_0\right]=0, 
	\end{equation}
	or in the format of virtual displacements as
	\begin{equation}
	\label{nholdiffskate}
	\Bigl[\delta X_0+\delta \zeta_l\Bigr]\cos\left[\theta(l)+\beta_0\right]+\delta Y(l)\sin\left[\theta(l)+\beta_0\right]=0,
	\end{equation}
	where 
	\begin{equation}
	\label{kinem2}
	\delta \zeta_l=\delta\left(\int_0^l \cos \theta(\varsigma,t) \text{d}\varsigma\right);
	\end{equation}
	
	\item \textit{V - the \lq violin bow' constraint}, expressed by equation (\ref{numerodue2}), which is rewritten as 
	\begin{equation}
	\label{vbowconstraint}
	\begin{split}
	\dot X(l)\cos\beta_0+\dot Y(l)\sin\beta_0+&\dot\theta(l)\Bigl[Y(l)\cos\beta_0 \\
	&+[l-X(l)]\sin\beta_0\Bigr]=0,
	\end{split}
	\end{equation}
	or, equivalently, in terms of virtual displacements as
	\begin{equation}
	\begin{split}
	\label{nholdiffvbow}
	\Bigl[\delta X_0+&\delta \zeta_l\Bigr]\cos\beta_0+\delta Y(l)\sin\beta_0\\
	&+\delta\theta(l)\Bigl[Y(l)\cos\beta_0+[l-X(l)]\sin\beta_0\Bigr]=0.
	\end{split}
	\end{equation}

\end{itemize}

As a secondary effect, the device realizing the non-holonomic constraint has a mass $M_l$ and rotational inertia $I_{r,l}$, which may be non-negligible and therefore are both assumed to act at the rod's end, $\textsf{s}=l$. 

\subsection{Dissipative effects} 
Different  sources of external dissipation are considered through the following linear damping coefficients:
\begin{itemize}
	\item $c_{e}$ and $c_{t,l}$ modelling \textit{external translational damping} respectively distributed on the rod and concentrated on the non-holonomic device (both 
	dissipation sources correspond to viscous forces, provided for instance by the air drag during motion);
	\item $c_{r,l}$ - \textit{rotational damping produced at the non-holonomic constraint} when pivoted about $\boldsymbol{e}_3$. 
\end{itemize}
Following \cite{bigonikiribook}, dissipation from \textit{internal damping} is introduced by means of the linear visco-elastic constitutive law $\sigma=E\,\epsilon+\Lambda \, \dot\epsilon$, relating the longitudinal stress $\sigma$ to the strain $\epsilon$ and its rate, respectively through the Young modulus $E$ and the  viscosity parameter $\Lambda$. Assuming a linear strain distribution ($\epsilon=y\,\theta'$, where $y$ is the distance from the neutral axis
and a dash \lq $\,\, \,' \,\,$ ' denotes differentiation with respect to the curvilinear coordinate $\textsf{s}$), integration over the rod's cross section  provides the internal bending moment $\mathcal{M}$ as
\begin{equation}
\begin{split}
\mathcal{M}(\textsf{s},t)=B\,\theta'(\textsf{s},t)+D\,\dot\theta'(\textsf{s},t)\,,
\end{split}
\end{equation}
where  $B=EJ$ and $D=\Lambda J$ are respectively 
the bending stiffness and the viscous damping coefficient associated to bending, with $J=\int_A y^2 \text{d}A$ being the second moment of inertia of the cross section.

\section{Equations of motion}

\subsection{The elastic rod}

Disregarding possible self-contact for the rod, its equations of motion can be obtained by considering the total potential energy $\mathcal{V}$ given by  the sum of elastic energy (stored in the rod and, when present, in the spring) and the potential of the load (when present) 
\begin{equation}
\label{Celasticenergy}
\mathcal{V}=\Xi(X_0)+\frac{B}{2} \int_0^l \theta'(\textsf{s},t)^2 \text{d}\textsf{s} ,
\end{equation}
where $\Xi$ is given either by equation (\ref{plof1}) or (\ref{plof2}). 

Considering the different inertial contributions to the system, the kinetic energy $\mathcal{T}$ is given by 
\begin{equation}
\label{Ckinetic}
\begin{split}
\mathcal{T}=&\frac{1}{2} I_{r,l} \dot{\theta}(l,t)^2  +\frac{1}{2} M_l \left(\dot{X}(l,t)^2+\dot{Y}(l,t)^2\right)\\&+\frac{1}{2} M_X \dot{X}_0(t)^2+\frac{1}{2}\rho\int_0^l\left(\dot{X}(\textsf{s},t)^2+\dot{Y}(\textsf{s},t)^2\right)\text{d}\textsf{s},
\end{split}
\end{equation}
where $\rho$ is the uniform  mass density. 

It follows that the Lagrangian $\mathcal{L}$ of the system is
\begin{equation}
\begin{split}
\label{Clag}
\mathcal{L}=\int_{0}^{l} \mathscr{L} \text{d}\textsf{s},
\end{split}
\end{equation}
where 
\begin{equation}
\label{Clagdens}
\begin{split}
\mathscr{L}=&\frac{M_X}{2 l} \dot{X}_0(t)^2+\frac{I_{r,l}}{2 l} \dot{\theta}(l,t)^2 +\frac{M_l}{2l} \left(\dot{X}(l,t)^2+\dot{Y}(l,t)^2\right)\\&+\frac{\rho }{2}\left(\dot{X}(\textsf{s},t)^2+\dot{Y}(\textsf{s},t)^2\right)-\frac{\Xi(X_0)}{l}-\frac{1}{2} B \theta'(\textsf{s},t)^2\\
&+\frac{M_0}{l}\theta(0)+R_X(X'-\cos \theta)+R_Y(Y'-\sin \theta)\,,
\end{split}
\end{equation}
in which the time derivative $\dot{X}$ and $\dot{Y}$  of the position functions $X$ and $Y$ can be obtained by differentiating the Eqs.(\ref{kinem}) and the 
three Lagrangian multipliers $R_X(\textsf{s},t)$, $R_Y(\textsf{s},t)$ and $M_0(t)$ are associated to the constraints of inextensibility (from the derivative with respect to $\textsf{s}$ of Eqs.(\ref{kinem})) and of null rotation of the initial rod's end, respectively. 

The equations of motion for the rod 
(without keeping into account for the moment non-holonomic constraints and dissipative effects) can be obtained by means of the virtual work of the conservative forces acting on the system and the vanishing of the first variation $\delta\mathcal{A}$ of the \lq action' integral $\mathcal{A}$, 
\begin{equation}
\label{deltaa}
\delta\mathcal{A} = 0, ~~~ \mbox{where} ~~~ \mathcal{A}=\int_{t_0}^{t_1} \int_{0}^{l} \mathscr{L} \text{d}\textsf{s} \,dt ,
\end{equation}
and $t_0$ and $t_1$ are two arbitrary time instants. 
To enforce equation (\ref{deltaa}), the following primary functions are needed
\begin{equation}
\begin{split}
\label{Cdofs}
\textbf{\textit{w}}=\{X_0,\,X,\,Y,\theta,\,X(l),\,Y(l),\theta(l),\,R_X,\,R_Y,\,M_0\},
\end{split}
\end{equation}
whose variations $\delta \textbf{\textit{w}}$ are subject to the boundary conditions at the rod's ends ($\textsf{s}=0$ and $\textsf{s}=l$) 
and at the two time instants ($t_0$ and $t_1$), 
\begin{equation}
\begin{split}
\label{nullvarstime}
&\delta\textbf{\textit{w}}\bigl|_{t_0}=\delta\textbf{\textit{w}}\bigl|_{t_1}=\textbf{0},\quad\delta\theta(0)=\delta Y(0)=0,\\
& \delta X(0)=\delta X_0,\quad \delta X(l)=\delta X_0+\delta \zeta_l,
\end{split}
\end{equation}
where $\delta \zeta_l$ is specified by Eq.(\ref{kinem2}).

\subsection{Equations of motion for the non-holonomic systems with dissipative effects}

The equation 
(\ref{deltaa}) introduced in the previous section hold for any elastic rod with the left end connected to a freely-sliding clamp, so that in these equations the non-holonomic constraints and dissipations do not play, for the moment, any role. Their role is defined as follows.

Dissipation can be introduced in the D'Alembert-Lagrange equations by adding to $\delta\mathcal{A}$ the virtual work of damping forces through the Rayleigh dissipation function $\mathcal{F}_d$, built from the 4 considered 
viscous sources as 
\begin{equation}
\label{Cdissipfun}
\mathcal{F}_d=\int_0^l{\mathscr{F}_d}\,\text{d}\textsf{s}, 
\end{equation}
where
\begin{equation}
\begin{split}
\label{Cdissipfun2}
\mathscr{F}_d=&\frac{c_{e}}{2}  (\dot{X}(\textsf{s},t)^2+\dot{Y}(\textsf{s},t)^2) 
+ \frac{c_{r,l}}{2l}  \dot{\theta}(l,t)^2\\&+\frac{D}{2} \dot{\theta}'(\textsf{s},t)^2  + \frac{c_{t,l}}{2l}  \left(\dot{X}(l,t)^2+\dot{Y}(l,t)^2\right). 
\end{split}
\end{equation}

The non-holonomic constraint, Eq.(\ref{nholdiffskate}) or (\ref{nholdiffvbow}), can be introduced in the D'Alembert-Lagrange equations by means of the Lagrangian multiplier $p(t)$, representing the reaction force (positive when compressive) transmitted by the non-holonomic constraint to the structure. By keeping into consideration also viscosity, the complete equations become
\begin{equation}
\begin{split}
&\delta\mathcal{A}-\int_{t_0}^{t_1} 
\int_{0}^{l} \sum_i\left(\frac{\partial\mathscr{F}_d}{\partial \dot \alpha_i}\delta \alpha_i\right) \,\text{d}\textsf{s}\,\text{d}t=\\
&\int_{t_0}^{t_1} p(t)\Bigl[\Bigl(\delta X_0+\delta \zeta(l)\Bigr)\cos\Theta+\delta Y(l)\sin\Theta+\delta\theta(l)\,\mathsf{A}\Bigr]\text{d}t,
\end{split}
\end{equation}
where $\{\alpha_i\}=\{X,\,Y,\,\theta',\,\theta(l),\,X(l),\,Y(l)\}$ (its rate and variations are
respectively 
$\dot{\alpha}_i$ and $\delta \alpha_i$)
and
\begin{equation}
\label{tettoneA}
\begin{split}
&\Bigl\{\Theta,\,\mathsf{A}\Bigr\}=\\
&\left\{
\begin{array}{ll}
\Bigl\{\theta(l)+\beta_0,\,0\Bigr\}\qquad\qquad\qquad & \text{\lq skate',}\\[3mm]
\Bigl\{\beta_0,\,Y(l)\cos\beta_0+[l-X(l)]\sin\beta_0\Bigr\} &\text{\lq violin bow'.}
\end{array}
\right.
\end{split}
\end{equation}

Invoking arbitrariness of the independent variations of the parameters and of the time instants $t_0$ and $t_1$, the equations of motion can be finally obtained. Arbitrariness of variations in the Lagrangian multipliers leads to
\begin{equation}
\label{ELeqhol}
X'=\cos\theta,\quad Y'=\sin\theta,\quad\theta(0,t)=0,
\end{equation}
while the field equations governing functions $X$, $Y$ and $\theta$ are 
\begin{equation}
\label{ELeqXY}
\rho \ddot X+c_e \dot X+R_X'=0,\qquad \rho \ddot Y+c_e \dot Y+R_Y'=0,
\end{equation}
and
\begin{equation}
\label{ELeqtheta}
B\theta''+D\dot \theta''+R_X \sin\theta-R_Y \cos\theta=0,
\end{equation}
respectively. The coordinate $X_0$ of the moving clamp satisfies
\begin{equation}
\label{ELeqX0}
\begin{split}
&M_X \ddot X_0+\frac{\text{d}\,\Xi(X_0)}{\text{d}\,X_0}+M_l \ddot X(l)+c_{t,l}\dot X(l)-R_X(l)\\
&+R_X(0)+p \cos\Theta=0. 
\end{split}
\end{equation}
Finally, the equations of motion are complemented by the contributions governing $X(l)$, $Y(l)$
\begin{equation}
\label{ELeqBCs}
\begin{split}
&M_l \ddot X(l)+c_{t,l} \dot X(l)-R_X(l)+p \cos\Theta=0,\\
&M_l \ddot Y(l)+c_{t,l} \dot Y(l)-R_Y(l)+p \sin\Theta=0,
\end{split}
\end{equation}
the rotation of the end of the rod $\theta(l)$ 
\begin{equation}
\label{ELeqBCtheta}
I_{r,l}\ddot\theta(l)+c_{r,l}\dot\theta(l)+B\theta'(l)+D\dot\theta'(l)+p\,\mathsf{A}=0 ,
\end{equation}
and by the non-holonomic constraint
\begin{equation}
\label{ELeqNH}
\dot X(l)\cos\Theta+\dot Y(l)\sin\Theta+\dot \theta(l)\,\mathsf{A}=0.
\end{equation}
The number of governing equations can be reduced through integration of Eqs.(\ref{ELeqXY}) to provide expressions for 
the Lagrangian multipliers $R_X(\textsf{s},t)$ and $R_Y(\textsf{s},t)$ 
\begin{equation}
\begin{array}{l}
R_X=R_X(l)+\int_\textsf{s}^l \left(\rho \ddot X+c_e \dot X\right)\,\text{d}\varsigma , \\[2mm]
R_Y=R_Y(l)+\int_\textsf{s}^l \left(\rho \ddot Y+c_e \dot Y\right)\,\text{d}\varsigma ,
\end{array}
\end{equation}
which represent the space evolution of the components of the total internal force projected along the horizontal $X$ and vertical $Y$ axes, respectively. By means of equations (\ref{ELeqBCs}), the terms $R_X(l,t)$ and $R_Y(l,t)$ can be made explicit, 
\begin{equation}
\label{gnocca}
\begin{array}{l}
R_X=M_l \ddot X(l)+c_{t,l} \dot X(l)+p \cos\Theta+\int_\textsf{s}^l \left(\rho \ddot X+c_e \dot X\right)\,\text{d}\varsigma,\\[2mm]
R_Y=M_l \ddot Y(l)+c_{t,l} \dot Y(l)+p \sin\Theta+\int_\textsf{s}^l \left(\rho \ddot Y+c_e \dot Y\right)\,\text{d}\varsigma.
\end{array}
\end{equation}
Eqs.(\ref{gnocca}) therefore represent the balance of linear momentum along the $X$ and $Y$ directions for the portion of rod comprised between the coordinates $\textsf{s}$ and $l$. 

The substitution of Eqs.(\ref{gnocca}) into Eq.(\ref{ELeqtheta}) yields the following expression for the rotation field
\begin{equation}
\label{eqtheta}
\begin{split}
&B\theta''+D\dot \theta''+p \sin(\theta-\Theta)+\sin\theta\Bigl(M_l \ddot X(l)+c_{t,l}\dot X(l)\\
&+ \int_\textsf{s}^l \left(\rho \ddot X+c_e \dot X\right)\,\text{d}\varsigma\Bigr)-\cos\theta\Bigl(M_l \ddot Y(l)+c_{t,l} \dot Y(l)\\
&+\int_\textsf{s}^l \left(\rho \ddot Y+c_e \dot Y\Bigr)\,\text{d}\varsigma\right)=0,
\end{split}
\end{equation}
while the coordinate $X_0$ can be expressed by
\begin{equation}
\label{eqX0}
\begin{split}
&M_X \ddot X_0+\frac{\text{d}\,\Xi(X_0)}{\text{d}\,X_0}+M_l \ddot X(l)+c_{t,l}\dot X(l)+p \cos\Theta\\
&+\int_\textsf{0}^l \left(\rho \ddot X+c_e \dot X\right)\,\text{d}\varsigma=0.
\end{split}
\end{equation}

Therefore, Eqs.(\ref{eqtheta}), (\ref{eqX0}), (\ref{ELeqNH}),  (\ref{ELeqhol})$_1$ and (\ref{ELeqhol})$_2$ represent a system of 5 equations in the 5 unknown functions $\{\theta,\,X_0,\,p,\,X,\,Y\}$, complemented by Eqs.(\ref{ELeqhol})$_3$ and (\ref{ELeqBCtheta}) expressing the boundary conditions for $\theta$.

\section{Euler's elastica with non-holonomic constraints}
\label{QSelasticanhol}

Neglecting the inertial and damping terms in the aforementioned equations, the equations governing the Euler's elastica are recovered, subject to a \lq skate' or a \lq violin bow' non-holonomic constraint. Under the quasi-static assumption, the Eqs.(\ref{ELeqtheta}), (\ref{ELeqhol}), (\ref{gnocca}) and (\ref{eqX0}) reduce to the following system governing the quasi-static configuration (superscript $^{\mbox{\tiny{QS}}}$) 
\begin{subequations}
	\label{StaticeqC}
	\begin{empheq}[left=\empheqlbrace]{align}
	&B\theta''^{\mbox{\tiny{QS}}}+R_{X}^{\mbox{\tiny{QS}}} \sin\theta^{\mbox{\tiny{QS}}}-R_{Y}^{\mbox{\tiny{QS}}} \cos\theta^{\mbox{\tiny{QS}}}=0\\[2mm]
	&X'^{\mbox{\tiny{QS}}}=\cos\theta^{\mbox{\tiny{QS}}}\\[2mm]
	&Y'^{\mbox{\tiny{QS}}}=\sin\theta^{\mbox{\tiny{QS}}}\\[2mm]
	&R_{X}^{\mbox{\tiny{QS}}}=p^{\mbox{\tiny{QS}}} \cos\Theta^{\mbox{\tiny{QS}}}=\text{const}\\[2mm]
	&R_{Y}^{\mbox{\tiny{QS}}}=p^{\mbox{\tiny{QS}}} \sin\Theta^{\mbox{\tiny{QS}}}=\text{const}\\[2mm]
	&\frac{\text{d}\,\Xi(X_0^{\mbox{\tiny{QS}}})}{\text{d}\,X_0^{\mbox{\tiny{QS}}}}+p^{\mbox{\tiny{QS}}} \cos\Theta^{\mbox{\tiny{QS}}}=0\,,
	\end{empheq}
\end{subequations}
where the equation for the non-holonomic constraint (\ref{ELeqNH}) is automatically satisfied. Moreover, by considering Eqs.(\ref{ELeqhol})$_3$ and (\ref{ELeqBCtheta}), the aforementioned system is complemented by the following boundary conditions on the rotation field $\theta^{\mbox{\tiny{QS}}}$
\begin{equation}
\label{BCQS}
\theta^{\mbox{\tiny{QS}}}(0)=0,\quad B\theta'^{\mbox{\tiny{QS}}}(l)+p^{\mbox{\tiny{QS}}}\mathsf{A}^{\mbox{\tiny{QS}}}=0
\end{equation}
while the condition $Y^{\mbox{\tiny{QS}}}(0)=0$ holds true.

Similarly to the case of discrete systems \cite{cazzollinhol}, the number of unknowns $\{\theta^{\mbox{\tiny{QS}}},\,R^{\mbox{\tiny{QS}}}_X,\,R^{\mbox{\tiny{QS}}}_Y,\,X^{\mbox{\tiny{QS}}},\,Y^{\mbox{\tiny{QS}}},\,X^{\mbox{\tiny{QS}}}_0,\,p^{\mbox{\tiny{QS}}}\}$ is equal to 7, thus exceeding the number of Eqs.(\ref{StaticeqC}) of a factor 1.
This reflects the fact that the quasi-static solution is represented by a one-dimensional \textit{manifold of equilibrium states} \cite{neimark} that will be 
shown below to be parametrized through the free coordinate $\theta^{\mbox{\tiny{QS}}}(l)$. 

Equilibrium configuration can be found in terms of Jacobi elliptic functions for the inflectional elastica  \cite{cazzolli}, by fixing the following four parameters:
\begin{itemize}
	\item[-] the left end rotation of the rod $\theta^{\mbox{\tiny{QS}}}(0)=0$, Eq.~(\ref{BCQS})$_1$;
	\item[-] the right end rotation of the rod $\theta^{\mbox{\tiny{QS}}}(l)$;
	\item[-] the inclination  $\beta=\Theta^{\mbox{\tiny{QS}}}$ of $p^{\mbox{\tiny{QS}}}$ w.r.t. the $X$-axis; 
	\item[-] 
	the characteristic parameter of the elastica $\eta$, representing the inclination of the tangent to the elastica at an inflection point with respect to the line 
	connecting all inflections.
\end{itemize}
In the following, the parameter $\eta$ is obtained for the rod subject to the \lq skate' and the \lq violin bow' constraints as a 
function of the end rotation $\theta^{\mbox{\tiny{QS}}}(l)$.

\subsection{Statics of the rod for \lq skate' constraint}
For the \lq skate' constraint, the parameter $\beta$ is the \lq skate' inclination with respect to the $X$ axis,  $\beta=\Theta^{\mbox{\tiny{QS}}}=\theta^{\mbox{\tiny{QS}}}(l)+\beta_0$. Furthermore, because $\mathsf{A}^{\mbox{\tiny{QS}}}=0$, by exploiting the analytical expression (\ref{BCQS})$_2$ of vanishing curvature at the final end $\theta'^{\mbox{\tiny{QS}}}(l)=0$ it follows that $\eta=\sin(|\beta_0|/2)$.\footnote{A null curvature at the coordinate $\textsf{s}=l$ implies 
	\begin{equation*}
	\cos\left\{\arcsin{\left[\frac{1}{\eta}\sin\left(\frac{-\beta_0}{2}\right)\right]}\right\}=0,\quad\rightarrow\quad \sqrt{1-\frac{1}{\eta^2}\sin^2\left(\frac{\beta_0}{2}\right)}=0.
	\end{equation*}}

Because the elastica provides the non-holonomic reaction force $p^{\mbox{\tiny{QS}}}=|R|$ as a function of the end's rotation $\theta^{\mbox{\tiny{QS}}}(l)$, this angle represents the only unknown needed to define the equilibrium configuration. Indeed, the clamp position $X_0^{\mbox{\tiny{QS}}}$ can be solved from equation (\ref{StaticeqC}f) as a function of $\Delta$ (for loading A), or of $F$ (for loading B), and $\theta^{\mbox{\tiny{QS}}}(l)$.

It follows that the rotation $\theta^{\mbox{\tiny{QS}}}(l)$ plays the role of a free coordinate of the system (\ref{StaticeqC}), so that the existence of the inflectional elastica (related to real values of parameter $\omega_0$ in \cite{cazzolli}) provides the following manifold of equilibrium states for the elastica subject to a \lq skate' non-holonomic constraint
\begin{equation}
\label{manrod}
\mathcal{D}:=\left\{\theta^{\mbox{\tiny{QS}}}(l)\in \mathbb{R}\,\,:\,\,\left|\dfrac{\sin \left(\dfrac{\theta^{\mbox{\tiny{QS}}}(l)+\beta_0}{2} \right)}{\sin\dfrac{|\beta_0|}{2}}\right|\leq 1\right\}.
\end{equation}

\begin{figure}[!htb]
	\begin{center}
		\includegraphics[width=3.2 in]{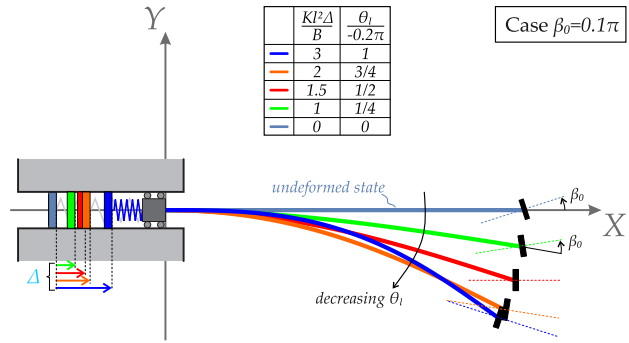}
	\end{center}
	\caption{\footnotesize{Five different equilibrium configurations for an elastic rod subject to the non-holonomic \lq skate' constraint inclined at $\beta_0=0.1\pi$. The five deformed configurations share the same position of the sliding clamp, $X_0^{\mbox{\tiny{QS}}}=0$, but differ in the value of the end's rotation $\theta^{\mbox{\tiny{QS}}}(l) $ and the corresponding force in the spring (dimensionless values $Kl^2\,\Delta/B$ listed in the legend).}}
	\label{fig_continuumQS}
\end{figure}

As examples of multiple solutions, five quasi-static configurations selected within the manifold of equilibrium states are reported in Fig.\ref{fig_continuumQS} for a \lq skate' inclination $\beta_0=0.1\pi$. The configurations share a null position of the sliding clamp, $X_0^{\mbox{\tiny{QS}}}=0$, but differ in the end's rotation value $\theta^{\mbox{\tiny{QS}}}(l)$ and in the imposed displacement $\Delta$ of the spring. These configurations are picked within the set defined by equation (\ref{manrod}) which for the considered \lq skate' inclination reduces to $\theta^{\mbox{\tiny{QS}}}(l)\in\left[-0.2,0\right]\pi$.

The condition $\eta=\sin(|\beta_0|/2)$ shows that the trivial (straight) solution  is unique for a \lq skate' aligned with the rod's end, $\beta_0=0$. Indeed, the manifold domain (\ref{manrod}) reduces to the condition $\theta^{\mbox{\tiny{QS}}}(l)=0$, so that $\theta^{\mbox{\tiny{QS}}}(\textsf{s})=0$ for every $\textsf{s}\in\left[0,\,l\right]$.

\subsection{Statics of the rod for \lq violin bow' constraint}

For \lq violin bow' constraint, the parameter $\beta$ is the inclination of the rotating cylinder, $\beta=\beta_0$. 
The elastica parameter $\eta$ is provided through the following nonlinear relation, obtained from Eq.(\ref{BCQS})$_2$ by considering Eq.(\ref{tettoneA}) 
and involving the end rotation $\theta^{\mbox{\tiny{QS}}}(l)$ 
\begin{equation}
\label{etatheta}
2\eta\cos\omega_0=-\left[F(\omega_l,\eta)-F(\omega_0,\eta)\right]\sin\beta_0,
\end{equation}
where $F(\sigma,\kappa)$ is the \textit{incomplete elliptic integral of the first kind} \cite{cazzolli} and for $m$ inflection points
\begin{equation*}
\begin{split}
&\omega_0=\arcsin\left[\frac{1}{\eta}\sin\left(\frac{-\beta_0}{2}\right)\right],\\
&\omega_l=(-1)^m\arcsin\left[\frac{1}{\eta}\sin\left(\frac{\theta^{\mbox{\tiny{QS}}}(l)-\beta_0}{2}\right)\right]\pm m\,\pi . 
\end{split}
\end{equation*}
As for the \lq skate' constraint, now the end's rotation $\theta^{\mbox{\tiny{QS}}}(l)$ of the rod plays the role of free coordinate defining a one-dimensional manifold of equilibrium states.
The uniqueness of the trivial solution at null inclination $\beta_0=0$ is proven by Eq.(\ref{etatheta}), which reduces to $\eta=0$ and leads to $\theta^{\mbox{\tiny{QS}}}(\textsf{s})=0$ for every $\textsf{s}\in\left[0,\,l\right]$.

\section{Linearized dynamics and instabilities. Case $\beta_0=0$}

In the next Sections, the linearized equations of motion for the considered non-holonomic systems are obtained and particularized for $\beta_0=0$. This condition represents a perfectly-aligned non-holonomic constraint, for which the sliding direction is perpendicular to the tangent vector at the final end of the rod $\textsf{s}=l$ for the \lq skate' constraint and for which the inclination of the rolling cylinder taken with respect to the horizontal is null for the \lq violin bow' constraint. Moreover, as sentenced in the previous Section, the condition $\beta_0=0$ corresponds to the existence of a unique quasi-static solution represented by the trivial one.

The linearized equations of motion around the trivial quasi-static solution, $\theta^{\mbox{\tiny{QS}}}=0$ and $X_0^{\mbox{\tiny{QS}}}=0$, are obtained by introducing an arbitrary small real parameter $\epsilon$ relating the  fields and quantities to their unperturbed (quasi-static, superscript $^{\mbox{\tiny{QS}}}$) and perturbed contributions (superimposed \lq$\,\,\hat{ }\,\,$')
\begin{equation}
\theta=\epsilon\,\hat\theta,\,\qquad X_0=\epsilon\,\hat X_0,\,\qquad p=p^{\mbox{\tiny{QS}}}+\epsilon\,\hat p,
\end{equation}
and by assuming fixed displacement $\Delta$. The reaction force at equilibrium $p^{\mbox{\tiny{QS}}}$ can be evaluated through Eq.(\ref{StaticeqC}f) in the specific case $\theta(l)=\beta_0=0$ thus obtaining

\begin{equation}
\label{pqsloads}
p^{\mbox{\tiny{QS}}}=\left\{
\begin{array}{ll}
K(\Delta-X_0^{\mbox{\tiny{QS}}})\qquad\qquad & \text{for elastic device A,}\\[3mm]
F &\text{for dead loading B.}
\end{array}
\right.
\end{equation}

The linearization of the equations of motion (\ref{eqtheta}) and (\ref{eqX0}) and of the boundary conditions equations (\ref{ELeqhol})$_3$ and (\ref{ELeqBCtheta}) is performed by considering the first-order term in the expansion in $\epsilon$. For simplicity, the symbol \lq$\,\,\hat{ }\,\,$' is removed henceforth so that quantities $\left\{\theta,\,X_0,\,p\right\}$ will denote perturbations of the related scalar fields.

Exploiting Eqs.(\ref{kinem}) for $X$ and $Y$, the equation of motion (\ref{eqtheta}) is linearized as 
\begin{equation}
\label{lintheta1}
\begin{split}
&B\theta''+D\dot \theta''-\int_\textsf{s}^l \left(\rho \int_0^\varsigma \ddot \theta \text{d}\sigma+c_e \int_0^\varsigma \dot \theta \text{d}\sigma\right)\,\text{d}\varsigma\\
&+p^{\mbox{\tiny{QS}}} \left[\theta-\Gamma\,\theta(l)\right]-M_l\int_0^l \ddot \theta \text{d}\textsf{s}-c_{t,l}\int_0^l \dot \theta \text{d}\textsf{s}=0\,,
\end{split}
\end{equation}
where the reaction force of the non-holonomic constraint is present both with its unperturbed (\lq pre-stress') part $p^{\mbox{\tiny{QS}}}$ at the equilibrium and its perturbation $p$, and $\Gamma = 1$ ($\Gamma = 0$) for \lq skate' (for \lq violin bow') constraint. 

The linearization of Eq.(\ref{ELeqhol})$_2$ leads to the condition $Y'=\theta$, so that a substitution in Eq.(\ref{lintheta1}) provides 
\begin{equation}
\label{lintheta05}
\begin{split}
&B \,Y'''+D\,\dot Y'''-\int_\textsf{s}^l \left(\rho \int_0^\varsigma \ddot Y' \text{d}\sigma+c_e \int_0^\varsigma \dot Y' \text{d}\sigma\right)\,\text{d}\varsigma\\
&+p^{\mbox{\tiny{QS}}} [Y'-\Gamma\,Y'(l)]-M_l\int_0^l \ddot Y' \text{d}\textsf{s}-c_{t,l}\int_0^l \dot Y' \text{d}\textsf{s}=0,
\end{split}
\end{equation}
which differentiated in $\textsf{s}$ yields 
\begin{equation}
\label{linY}
B \,Y''''+D\,\dot Y''''+ \rho \,\ddot Y +c_e\, \dot Y+p^{\mbox{\tiny{QS}}}\,Y''=0\,,
\end{equation}
where $p^{\mbox{\tiny{QS}}}$ is given by Eq.(\ref{pqsloads}). 
Eq.(\ref{linY}) is independent of $\Gamma$, so that it is valid for both types of non-holonomic constraints.

Due to the presence of fourth order spatial derivatives in the equation (\ref{linY}), a further boundary condition is introduced by particularising the differential equation (\ref{lintheta05}) at $\textsf{s}=l$, 
\begin{equation}
\begin{split}
\label{bcy1}
&B\, Y'''(l)+D\,\dot Y'''(l)+p^{\mbox{\tiny{QS}}} (1-\Gamma) Y'(l)-M_l\, \ddot Y(l) \\
&-c_{t,l}\, \dot Y(l)=0.
\end{split}
\end{equation}
The boundary condition (\ref{ELeqBCtheta}) can be linearized as
\begin{equation}
\label{bcy2}
\begin{split}
&B \,Y''(l)+D\,\dot Y''(l)+I_{r,l}\,\ddot Y'(l)+p^{\mbox{\tiny{QS}}} Y(l)(1-\Gamma)\\
&+c_{r,l}\,\dot Y'(l)=0\,,
\end{split}
\end{equation}
and complemented by $Y'(0,t)=Y(0,t)=0$. 

Because $\dot X_0=\dot X(l)$, both non-holonomic constraints, Eq.(\ref{ELeqNH}), are described by the same linearized equation
\begin{equation}
\label{linNH}
\dot X_0=0\quad\rightarrow \quad X_0(t)=X_0(0)\quad \forall \, t ,
\end{equation}
so that $X_0(0)=0$ is selected and the linearized version of the Eq.(\ref{eqX0}) gives a null perturbation in the reaction force, $p=0$, for every loading condition; 
therefore the reaction of the non-holonomic constraint remains constant in a first-order analysis, as it happens in the case of follower load.

Note that the type of constraint only affects  the boundary conditions (\ref{bcy1}) and (\ref{bcy2}) through $\Gamma$, while the loading device affects Eqs.(\ref{linY}), (\ref{bcy1}) and (\ref{bcy2}) through $p^{\mbox{\tiny{QS}}}$, see Eq.(\ref{pqsloads}).

The linearized equations governing the dynamics of the rod with the non-holonomic constraints, Eqs. (\ref{linY})--(\ref{bcy2}) plus $Y'(0,t)=Y(0,t)=0$, coincide for $\Gamma=1$ (for $\Gamma=0$) with the corresponding equations pertinent to the Beck column (to the Reut column), in its variant with viscosity and additional mass concentrated at the loaded end.  
Therefore, the models by Beck and Reut can be understood as simplified linear models for structures subject to non-holonomic constraints. Moreover, as noted in \cite{ruina} for a different non-holonomic system, the linearized equations of dynamics, Eqs. (\ref{linY})--(\ref{bcy2}) plus $Y'(0,t)=Y(0,t)=0$, violate the conservation of the energy (as the Beck and Reut models do), even though the exact equations without dissipation crearly describe a fully-conservative system. This peculiarity of non-holonomic systems is a consequence of the mechanical equivalence between non-holonomic conditions and polygenic \cite{lanczos} forces that cannot be derived from a potential.

\subsection{Dimensionless formulation} 

It is instrumental to introduce the following dimensionless quantities
\begin{equation}
\begin{split}
& s=\frac{\textsf{s}}{l},\quad\tau=\frac{t}{T},\quad\tilde{Y}=\frac{Y}{l},\quad\tilde p^{\mbox{\tiny{QS}}}=\frac{p^{\mbox{\tiny{QS}}} l^2}{B},\\
&\tilde M_l=\frac{M_l}{\rho l},\quad\tilde I_{r,l}=\frac{I_{r,l}}{\rho l^3 },\quad\tilde c_e=\frac{c_e l^2}{\sqrt{\rho B}},\\
&\tilde{D}=\frac{D}{l^2\sqrt{\rho B}},\quad \tilde c_{t,l}=\frac{c_{t,l} l}{\sqrt{\rho B}},\quad\tilde c_{r,l}=\frac{c_{r,l}}{l\sqrt{\rho B}} ,
\end{split}
\end{equation}
where $T=l^2 \sqrt{\rho/B}$ is the characteristic time of the rod. 
By introducing the exponential solution $\tilde{Y}(s,\tau)=\Psi(s)e^{\Omega\tau}$, where $\Omega$ is the dimensionless eigenvalue and $\Psi(s)$ the generic eigenfunction  of the system, the following fourth order ordinary differential equation is obtained (\lq $\,\, '\,\,$'  stands for the derivative with respect to the dimensionless curvilinear coordinate $s$) 
\begin{equation}\label{ODE4}
(1+\tilde{D}\,\Omega)\,\Psi''''(s)+\tilde p^{\mbox{\tiny{QS}}}\,\Psi''(s) + (\Omega^2+\tilde c_e\, \Omega)\,\Psi(s)=0\,,
\end{equation}
complemented by the boundary conditions
\begin{subequations}
	\label{linnodim}
	\begin{empheq}[left=\empheqlbrace]{align}
	&(1+\tilde{D}\,\Omega)\,\Psi'''(1)-(\tilde M_l\, \Omega^2 +\tilde c_{t,l}\,\Omega) \,\Psi(1)\notag\\
	&+\tilde p^{\mbox{\tiny{QS}}}\,(1-\Gamma) \Psi'(1)=0\,,\\[2mm]
	&(1+\tilde{D}\,\Omega)\,\Psi''(1)+(\tilde I_{r,l}\,\Omega^2+\tilde c_{r,l}\,\Omega) \,\Psi'(1)\notag\\
	&+\tilde p^{\mbox{\tiny{QS}}}\,(1-\Gamma) \Psi(1)=0\,,\\[2mm]
	&\Psi'(0)=0\,,\\[2mm]
	&\Psi(0)=0.
	\end{empheq}
\end{subequations}

\subsection{Evaluation of the flutter and divergence loads}
The eigenfunction  $\Psi(s)$,  solution to Eq.(\ref{ODE4}), is given by
\begin{equation}
\label{Psisolflu}
\begin{split}
\Psi(s)=&A_1 \sin{\left(\lambda_1 s\right)}+A_2 \cos{\left(\lambda_1 s\right)}+A_3 \sinh{\left(\lambda_2 s\right)}\\
+&A_4 \cosh{\left(\lambda_2 s\right)}\,,
\end{split}
\end{equation}
where 
\begin{equation}
\lambda_{1,2}=\sqrt{
	\sqrt{\dfrac{\left(\tilde{p}^{\mbox{\tiny{QS}}}\right)^2-4\, \Omega  (\tilde{D} \, \Omega +1) (\tilde c_e+\Omega )}{4(\tilde{D} \, \Omega +1)^2}}
	\pm
	\frac{\tilde{p}^{\mbox{\tiny{QS}}}}{2\, (\tilde{D} \, \Omega +1)}
},
\end{equation}
Imposing the boundary conditions (\ref{linnodim}) to the general solution (\ref{Psisolflu}) leads to the  linear system
\begin{equation}
\label{matrixform}
\left\{\textbf{M}+\tilde p^{\mbox{\tiny{QS}}}\,(1-\Gamma)\textbf{P}\right\}\,
\textbf{A}=0
\end{equation}
where
\begin{equation}
\label{systemflu}
\begin{split}
&\textbf{M}=\left[
\begin{array}{cccc}
0 & 1 & 0 & 1\\
\lambda_1 & 0 & \lambda_2 & 0\\
a_{31} & a_{32} & a_{33} & a_{34}\\
a_{41} & a_{42} & a_{43} & a_{44}
\end{array}\right],\\\\
&\textbf{P}=\left[
\begin{array}{cccc}
0 & 0 & 0 & 0\\
0 & 0 & 0 & 0\\
\sin\lambda_1 & \cos\lambda_1 & \sinh\lambda_2 & \cosh\lambda_2\\
-\lambda_1\cos\lambda_1 & \lambda_1\sin\lambda_1 & -\lambda_2\cosh\lambda_2 & -\lambda_2\sinh\lambda_2
\end{array}\right]
\end{split}
\end{equation} 
and the following parameters have been introduced
\begin{equation}
\label{a3}
\left\{
\begin{array}{l}
a_{31}=\lambda_1 \cos {\lambda_1} \left(\tilde{c}_{r,l}\Omega+\tilde{I}_{r,l} \Omega^2 \right)-\lambda_1^2 (\tilde{D}  \Omega +1) \sin{\lambda_1},\\
a_{32}=-\lambda_1 \sin{\lambda_1}\left(\tilde{c}_{r,l}\Omega+\tilde{I}_{r,l} \Omega^2 \right)-\lambda_1^2 (\tilde{D}  \Omega +1) \cos {\lambda_1},\\
a_{33}=\lambda_2 \cosh{\lambda_2}\left(\tilde{c}_{r,l}\Omega+\tilde{I}_{r,l} \Omega^2 \right)+\lambda_2^2 (\tilde{D}  \Omega +1) \sinh{\lambda_2},\\
a_{34}=\lambda_2 \sinh{\lambda_2}\left(\tilde{c}_{r,l}\Omega+\tilde{I}_{r,l} \Omega^2 \right)+\lambda_2^2 (\tilde{D}  \Omega +1) \cosh{\lambda_2},\\
a_{41}=-\sin{\lambda_1} \left(\tilde{c}_{t,l} \Omega +\tilde{M}_{l} \Omega ^2\right)-\lambda_1^3 (\tilde{D}  \Omega +1) \cos {\lambda_1},\\
a_{42}=-\cos {\lambda_1} \left(\tilde{c}_{t,l} \Omega +\tilde{M}_{l} \Omega ^2\right)+\lambda_1^3 (\tilde{D}  \Omega +1) \sin{\lambda_1},\\
a_{43}=-\sinh{\lambda_2} \left(\tilde{c}_{t,l} \Omega +\tilde{M}_{l} \Omega ^2\right)+\lambda_2^3 (\tilde{D}  \Omega +1) \cosh{\lambda_2},\\
a_{44}=-\cosh{\lambda_2} \left(\tilde{c}_{t,l} \Omega +\tilde{M}_{l} \Omega ^2\right)+\lambda_2^3 (\tilde{D}  \Omega +1) \sinh{\lambda_2}.
\end{array}\right.
\end{equation}

When $\Gamma=0$ (and $\Gamma=1$), the following property holds
\begin{equation}
\label{detidentity}
\det{\Bigl[\textbf{M}+\tilde p^{\mbox{\tiny{QS}}}\,(1-\Gamma)\textbf{P}\Bigr]}=\det{\textbf{M}},
\end{equation}
so that the characteristic equation obtained by imposing the vanishing of the determinant (\ref{detidentity}) for the \lq skate' ($\Gamma=1$)  and for the \lq violin bow' ($\Gamma=0$) constraints coincide (which is an expected property, as this coincidence is also observed for the Beck and Reut columns) and are given by 
\begin{equation}
\label{Cdeterm}
\begin{split}
\sinh{\lambda_2} \Bigl[&(\lambda_1^2-\lambda_2^2) \sin{\lambda_1} \left(\lambda_1^2 \lambda_2^2 C_1^2-C_2 C_3\right)\\
&+\lambda_1 C_1 \left(\lambda_1^2+\lambda_2^2\right) \cos {\lambda_1} \left(\lambda_2^2 C_2-C_3\right)\Bigr]\\
+\lambda_2 \cosh{\lambda_2} \Bigl[&2 \lambda_1 \cos {\lambda_1} \left(\lambda_1^2 \lambda_2^2 C_1^2-C_2 C_3\right)\\
&+C_1 \left(\lambda_1^2+\lambda_2^2\right) \sin{\lambda_1} \left(\lambda_1^2 C_2+C_3\right)\Bigr]\\
+\lambda_1 \lambda_2 \Bigl[&C_1^2 \left(\lambda_1^4+\lambda_2^4\right)+2 C_2 C_3\Bigr]=0\,,
\end{split}
\end{equation}
where 
\begin{equation*}
C_1=1+\tilde{D}\,\Omega,\,\,\, C_2=\Omega\left(\tilde c_{r,l}+\Omega\,\tilde I_{r,l}\right),\,\,\, C_3=\Omega\left(\tilde c_{t,l}+\Omega\,\tilde M_l\right) .
\end{equation*}
The critical load of flutter instability can be obtained from Eq.(\ref{Cdeterm}) as the smallest dimensionless pre-stress $\tilde{p}^{\mbox{\tiny{QS}}}$ causing 
$\Omega$ to display a complex conjugate pair with positive real part, while the divergence load corresponds to the smallest value of $\tilde{p}^{\mbox{\tiny{QS}}}$ providing a 
real and positive eigenvalue $\Omega$. 

Because the characteristic equation (\ref{Cdeterm}) applies to the elastic rod connected to both the \lq skate' and \lq violin bow', flutter and divergence loads are the same for both structures and in turn coincide with the Beck and Reut columns.
This feature corresponds to an analogous property found for discrete columns made up of $N$ of rigid bars connected by visco-elastic hinges \cite{cazzollinhol}.

As a final remark, the trivial eigenvalue $\Omega=0$ is always a solution for Eq.(\ref{Cdeterm}), a feature typical of non-holonomic systems. This trivial solution can simply be omitted in the evaluation of the stability of configurations belonging to the manifold of equilibrium states \cite{neimark}. By omitting the trivial eigenvalue $\Omega=0$, Eq.(\ref{Cdeterm}) coincides with that holding for the aforementioned structures.

\subsection{Numerical examples: stability and Ziegler destabilization paradox}

\begin{figure*}[!h]
	\begin{center}
		\includegraphics[width=0.7\textwidth]{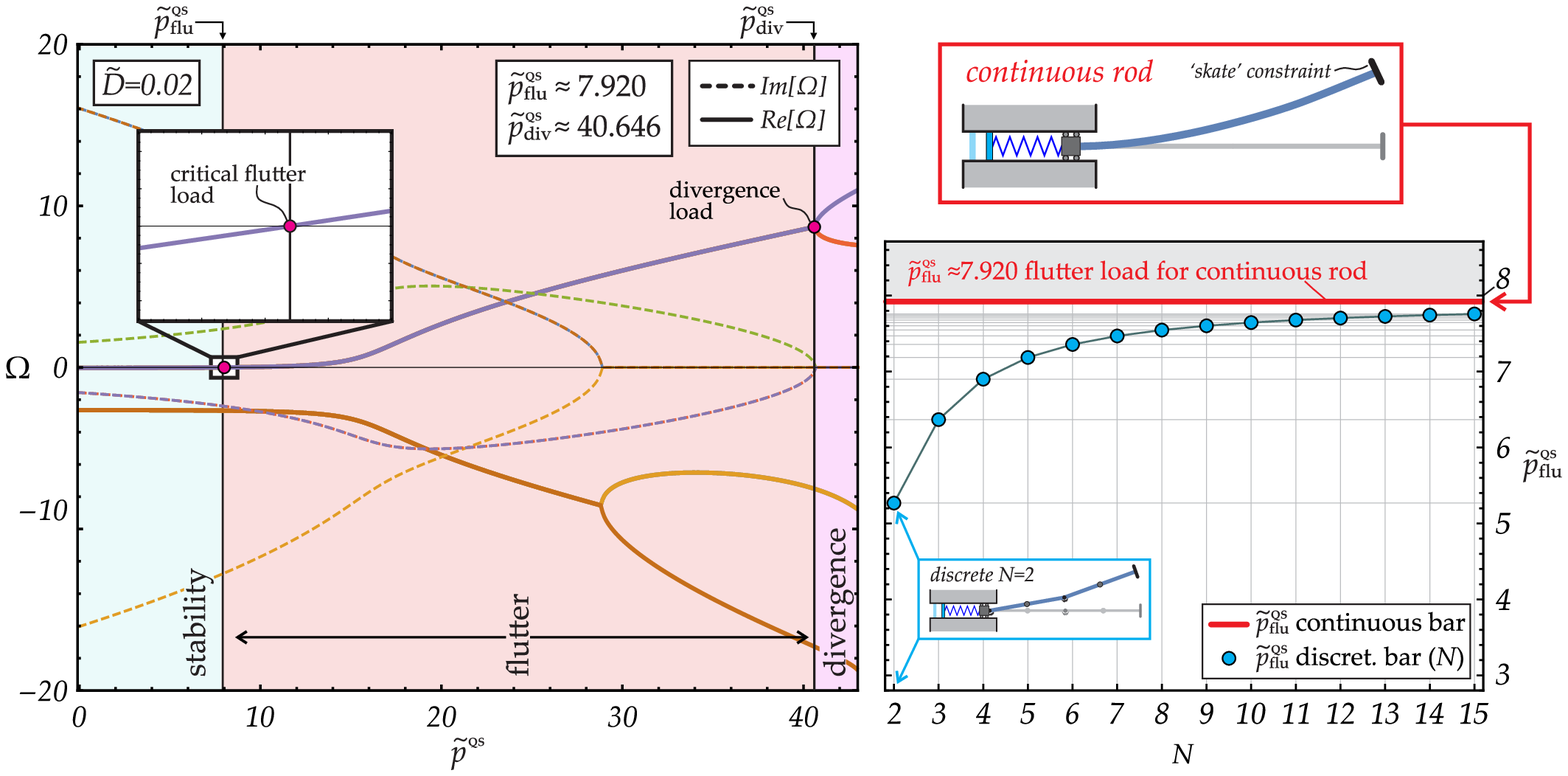}
	\end{center}
	\caption{\footnotesize{(Left) Branching of the real and imaginary parts of the eigenvalues $\Omega$ as functions of the dimensionless reaction force $\tilde{p}^{\mbox{\tiny{QS}}}$ for the non-holonomic \lq skate' constraint applied to a rod with $\tilde M_l=1$, $\tilde I_{r,l}=0$ and $\tilde D=0.02$. The flutter and divergence loads are $\tilde{p}^{\mbox{\tiny{QS}}}_{\text{flu}}\approx 7.920$ and $\tilde{p}^{\mbox{\tiny{QS}}}_{\text{div}}\approx 40.646$, respectively. (Right) Flutter load of the rod obtained as limit value at increasing number $N$ of elements for a chain of rigid bars connected through viscoelastic hinges \cite{cazzollinhol}.
	}}
	\label{fig_continuum_compare}
\end{figure*}

With reference to a rod having a mass ratio $\tilde M_l=1$, null rotational inertia $\tilde I_{r,l}=0$, and subject to an internal dissipation $\tilde D=0.02$ only ($\tilde c_e=\tilde c_{t,l}=\tilde c_{r,l}=0$), the branches of real and imaginary parts of the eigenvalues $\Omega$ 
are reported as a function of the pre-stress $\tilde{p}^{\mbox{\tiny{QS}}}$ in Fig.\ref{fig_continuum_compare} (left), showing the value of flutter and divergence loads, respectively $\tilde{p}^{\mbox{\tiny{QS}}}_{\text{flu}}\approx 7.920$ and $\tilde{p}^{\mbox{\tiny{QS}}}_{\text{div}}\approx 40.646$. 

Note that the flutter load  $\tilde{p}^{\mbox{\tiny{QS}}}_{\text{flu}}$ represents the limit value for a column made up of a large number $N$ of rigid bars (Fig.\ref{fig_continuum_compare}, right).\footnote{Using equations available in \cite{cazzollinhol} as Supplementary Material and following \cite{domokos}, the flutter load for the discrete counterpart of the continuous system can be evaluated  by considering the following  parameters of  internal rotational damping $c_i$ and of rotational stiffness $k_j$ ($j=1,...,N$) for a visco-elastic chain of $N$ rigid bars
	\begin{equation*}
	c_i=\frac{D}{l}N,\qquad k_1=\frac{2B}{l}N\,, \qquad k_j=\frac{B}{l}N,\qquad m_1=m_j=\frac{\rho l}{N},\qquad j=2,...,N \qquad 
	\end{equation*}
	and by assuming that the total length of the discrete system coincides with that of the continuous rod, $L=l$.
}

In agreement with the predictions for the Pfl\"uger column by Tommasini et al. \cite{tommy}, the \lq ideal' flutter and divergence loads are evaluated at null viscosities \lq from the beginning', for $\tilde M_l=1$ and $\tilde I_{r,l}=0$, to be $\mathcal{P}_0\approx 16.212$ and $\mathcal{D}_0\approx 34.465$, respectively. The flutter load in the limit of vanishing viscosities confirms the Ziegler's destabilization paradox for a rod subject to non-holonomic constraints. Indeed, such limit value of the flutter load is numerically found to be never higher than that evaluated for the same structure, but with null dissipation \lq from the beginning'. In particular, considering  the presence of a single damping source, the following values are found:
\begin{itemize}
	\item[-] \textit{Internal damping $\tilde D$} 
	\begin{equation*}
	\lim_{\tilde D\to 0}{\tilde{p}^{\mbox{\tiny{QS}}}_{\text{flu}}(\tilde D, \tilde c_e=\tilde c_{t,l}=\tilde c_{r,l}=0)}\approx 7.905 < \mathcal{P}_0\,;
	\end{equation*}
	\item[-] \textit{External damping $\tilde c_e$} 
	\begin{equation*}
	\lim_{\tilde c_e\to 0}{\tilde{p}^{\mbox{\tiny{QS}}}_{\text{flu}}(\tilde c_e, \tilde D=\tilde c_{t,l}=\tilde c_{r,l}=0)}\approx 16.122 < \mathcal{P}_0\,;
	\end{equation*}
	\item[-] \textit{Translational damping of the non-holonomic device $\tilde c_{t,l}$}
	\begin{equation*}
	\tilde{p}^{\mbox{\tiny{QS}}}_{\text{flu}}(\tilde c_{t,l}, \tilde D=\tilde c_e=\tilde c_{r,l}=0)\approx 16.052 < \mathcal{P}_0\,;
	\end{equation*}
	\item[-] \textit{Rotational damping of the non-holonomic device $\tilde c_{r,l}$} 
	\begin{equation*}
	\tilde{p}^{\mbox{\tiny{QS}}}_{\text{flu}}(\tilde c_{r,l}, \tilde D=\tilde c_e=\tilde c_{t,l}=\tilde c_{r,l}=0)\approx 5.219 < \mathcal{P}_0\,.
	\end{equation*}
	
\end{itemize}

Finally, it is worth highlighting that a \emph{viscosity-independent Ziegler destabilization  paradox} occurs for the continuous system as  observed for the discrete system \cite{cazzollinhol}. Indeed, when only one of the two damping sources connected to the non-holonomic device is considered, the flutter load $\tilde{p}^{\mbox{\tiny{QS}}}$ becomes independent of the damping coefficient (either $\tilde c_{t,l}$ or $\tilde c_{r,l}$), and smaller than the ideal value $\mathcal{P}_0$ for every amount of damping. 
\vspace{4mm}

\textbf{Acknowledgements}\\
 Financial  support  is  acknowledged  from  PITN-GA-2019-813424-INSPIRE, ARS01-01384-PROSCAN, and from the Italian Ministry of Education,  University and Research (MIUR) in the frame of the 'Departments of Excellence' grant L. 232/2016.



\begin{thebibliography}{99}
	
	\setlength{\itemsep}{-1.0mm}
	
	\bibitem{neimark}
	Neimark, Ju.I., Fufaev, N.A. (1972)
	\emph{Dynamics of Nonholonomic Systems}. Translations of Mathematical Monographs, V. 33.
	
	\bibitem{leipholz} Leipholz, H.H.E. (1980). Analysis of Nonconservative, Nonholonomic System. In: Theoretical and Applied Mechanics. Proceedings of the 15th International Congress on Theoretical and Applied Mechanics (ICTAM), Toronto (Canada), 17-23 August 1980 edited by F.P.J. Rimrott and B. Tabarrok, North-Holland Publishing Company, Amsterdam-New York-Oxford, 1-11.
	
	\bibitem{ruina}
	Meijaard, J.P., Papadopoulos, J.M., Ruina, A., Schwab, A.L. (2007) Linearized dynamics equations for the balance and steer of a bicycle: a benchmark and review. \emph{Proc. R. Soc. A}, vol. 463, 1955-1982.
	
	\bibitem{bottema} Bottema, O. (1949) On the small vibrations of non-holonomic systems.  \emph{Proceedings Koninklijke Nederlandse Akademie van Wetenschappen}, 1936, 848-850.
	
	\bibitem{beregi1} Beregi, S., Takacs, D., Stepan, G. (2019) Bifurcation analysis of wheel shimmy with non-smooth effects and time delay in the tyre?ground contact.  \emph{Nonlinear Dynamics}, 98, 841-858.
	
	\bibitem{ziegshimmy} Ziegler, H. (1938) Die Querschwingungen von Kraftwagenanh\"angern.  \emph{Ingenieur-Archiv}, 9, 96-108.
	
	\bibitem{ziegler} Ziegler, H. (1977) \emph{Principles of Structural Stability}. Birkh\"auser. 
	
	\bibitem{cazzollinhol}
	Cazzolli, A., Dal Corso, F., Bigoni, D. (2020).
	Non-holonomic constraints inducing flutter instability in structures under conservative loadings. \emph{J. Mech. Phys. Solids}, 138, 103919.
	
	\bibitem{beck}
	Beck, M., 1952. Die Knicklast des einseitig eingespannten, tangential gedr\"uckten Stabes. \emph{Z. Angew. Math. Phys.}, 3, 225.
	
	\bibitem{bigonimiss22} Bigoni, D. and Misseroni, D. and Tommasini, M. and Kirillov, O. and Noselli, G. (2018) Detecting singular weak-dissipation limit for flutter onset in reversible systems. \emph{Phys. Rev. E.} 97, 023003.
	
	\bibitem{bigonikiri} Bigoni, D., Kirillov, O.N., Misseroni, D., Noselli, G., Tommasini, M. (2018) Flutter and divergence instability in the Pfl\"uger column: Experimental evidence of the Ziegler destabilization paradox. \emph{J. Mech. Phys. Solids} 116, 99-116.
	
	\bibitem{detinko}
	Detinko, F.M.  (2003)
	Lumped damping and stability of Beck column with a tip mass. \emph{Int. J. Sol. Struct.} 40, 4479-4486.
	
	\bibitem{pfluger}
	Pfl\"uger, A. (1955). Zur Stabilit\"at des tangential gedr\"uckten Stabes. \emph{Z. Angew. Math. Mech.} 35 (5), 191.
	
	\bibitem{bolotin} Bolotin, V.V. (1963) \emph{Nonconservative Problems of the Theory of Elastic Stability}, Pergamon Press. 
	
	\bibitem{reut}
	Reut, V.I. (1939)  About the Theory of Elastic Stability. \emph{Proc. Odessa Inst. Civil and Communal Eng.}, No. 1.
	
	\bibitem{kirillov_1} Kirillov, O.N. (2005) A theory of the destabilization paradox in non-conservative systems. \emph{Acta Mech.} 174, 145-166.
	
	\bibitem{kirillov_2} Kirillov, O.N., Verhulst, F. (2010) Paradoxes of dissipation-induced destabilization or who opened Whitney's umbrella? \emph{Z. Angew. Math. Mech.} 90, 462-488. 
	
	\bibitem{abdulla}
	Abdullatif, M., Mukherjee, R., Hellum, A. (2018). Stabilizing and destabilizing effects of damping in non-conservative systems: Some new results.
	\emph{J. Sound and Vibration}, 413, 442-455.
	
	\bibitem{agostinelli} Agostinelli, D., Lucantonio, A., Noselli, G., DeSimone, A. (2020) Nutations in growing plant shoots: The role of elastic deformations due to gravity loading. \emph{J. Mech. Phys. Solids}, 136, 103702.
	
	\bibitem{phan}
	Phan, H., Shin, D., Heon Jeon, S., Young Kang, T., Han, P., Han Kim, G., Kook Kim, H., Kim, K., Hwang, Y., Won Hong, S. (2017). Aerodynamic and aeroelastic flutters driven triboelectric nanogenerators for harvesting broadband airflow energy. \emph{Nano Energy}, 33, 476-484.
	
	\bibitem{koiter} Koiter, W.T. (1996) Unrealistic follower forces. \emph{J. Sound and Vibration} 194, 636-638.
	
	\bibitem{elishakoff} Elishakoff, I. (2005) Controversy associated with the so-called ``follower force'': critical overview. \emph{Appl. Mech. Rev.} 58, 117-142.
	
	\bibitem{bigoninoselli} Bigoni, D., Noselli, G. (2011) Experimental evidence of flutter and divergence instabilities induced by dry friction. \emph{J. Mech. Phys. Solids} 59, 2208-2226.
	
	\bibitem{bigmiss}  Bigoni, D., Misseroni, D. (2020). Structures loaded with a force acting along a fixed straight line, or the "Reut's column problem". 
	\emph{J. Mech. Phys. Solids}  134, 103741.
	
	\bibitem{bigonikiribook} Bigoni, D. (2019). Flutter from Friction in Solids and Structures. In CISM Lecture Notes No. 586 \emph{Dynamic Stability and Bifurcation in Nonconservative Mechanics} (Ch. 1), edited by D. Bigoni and O.N. Kirillov, Springer. 
	
	\bibitem{cazzolli}
	Cazzolli, A., Dal Corso, F.  (2018)
	Snapping of elastic strips with controlled ends.
	\emph{Int. J. Sol. Struct.}, 162, 285-303.
	
	\bibitem{lanczos}
	Lanczos, C. (1952) \emph{The Variational Principles of Mechanics}. Oxford University Press.
	
	\bibitem{domokos}
	Domokos G. (2002)
	The Odd Stability of the Euler Beam. In: Seyranian A.P., Elishakoff I. (eds) \emph{Modern Problems of Structural Stability.} International Centre for Mechanical Sciences, vol 436. Springer, Vienna.
	
	\bibitem{tommy}
	Tommasini, M., Kirillov, O.N., Misseroni, D., Bigoni, D. (2016)
	The destabilizing effect of external damping: Singular flutter boundary for the Pflüger column with vanishing external dissipation. \emph{J. Mech. Phys. Sol.} 91, 204-215. 
	
\end{thebibliography}
\end{document}